\documentclass[letterpaper,twocolumn,10pt]{article}
\usepackage{usenix2019_v3}
\usepackage[english]{babel}
\usepackage{amsmath}
\usepackage{amssymb}
\usepackage{booktabs}
\usepackage{comment}
\usepackage{enumitem}
\usepackage{graphicx}
\usepackage{subfig}
\usepackage{tabularx}
\usepackage{tikz}
\usepackage{xspace}
\usepackage{longtable}
\usepackage{xurl}
\usepackage{array}
\usepackage{makecell}
\usepackage[subtle]{savetrees}
\setitemize{noitemsep,topsep=0pt,parsep=0pt,partopsep=0pt}

\usepackage{shortcuts}

\usepackage[framemethod=TikZ]{mdframed}
\mdfdefinestyle{Takeaway}{
    linecolor=black,
        outerlinewidth=1pt,
    roundcorner=0pt,
            innerrightmargin=20pt,
    innerleftmargin=20pt,
    backgroundcolor=gray!50!white}
\mdtheorem[style=Takeaway]{takeaway}{Takeaway}
\newcommand{\take}[1]{\begin{takeaway}#1\end{takeaway}}

\renewcommand{\paragraph}{\vspace{3pt}\noindent\textbf}
\begin{document}

\date{}
\title{A Deep Dive into VirusTotal:\\
	Characterizing and Clustering a Massive File Feed}

\author{
	{\rm Kevin van Liebergen}\\
	IMDEA Software Institute
	\and
	{\rm Juan Caballero}\\
	IMDEA Software Institute
		\and
	{\rm  Platon Kotzias}\\
	Norton Research Group
	\and
	{\rm Chris Gates}\\
	Norton Research Group
} 
\maketitle

\begin{abstract}

Online scanners analyze user-submitted files
with a large number of security tools
and provide access to the analysis results.
As the most popular online scanner, 
VirusTotal (VT) 
is often used for
determining if samples are malicious, 
labeling samples with their family, 
hunting for new threats, and 
collecting malware samples.
We analyze 328M VT reports for 235M samples
collected for one year through the VT file feed.
We use the reports to characterize the VT file feed in depth and 
compare it with the telemetry of a large \vendor.
We answer questions such as 
How diverse is the feed?
Does it allow building malware datasets for different filetypes?
How fresh are the samples it provides?
What is the distribution of malware families it sees?
Does that distribution really represent malware on user devices?

We then explore how to perform threat hunting at scale by 
investigating scalable approaches that can produce high purity clusters 
on the 235M feed samples.
We investigate three clustering approaches:
hierarchical agglomerative clustering (HAC),
a more scalable HAC variant for TLSH digests (HAC-T), and
a simple feature value grouping (FVG).
Our results show that HAC-T and FVG using selected features
produce high precision clusters on ground truth datasets.
However, only FVG scales to the daily influx of samples in the feed.
Moreover, FVG takes 15 hours to cluster the whole dataset of 235M samples.
Finally, we use the produced clusters for threat hunting,
namely for detecting 190K samples thought to be benign
(i.e., with zero detections) that
may really be malicious because they belong to 29K clusters where
most samples are detected as malicious.

\end{abstract}
\section{Introduction}
\label{sec:introduction}

Online scanners analyze samples (i.e., files)
submitted by users using a large number of security tools, and 
provide access to the analysis results through free and commercial APIs. 
The most popular online scanner is VirusTotal~\cite{vt} (VT), 
which acts a de-facto central malware sharing 
service for the security community.
According to its own last 7 days statistics
VT receives over 2M daily file submissions~\cite{vtStats} and 
has collected over 2.4 billion files since it started 
operating in 2004~\cite{vtPremiumServices}.
Each file analysis by VT produces a \emph{report}
containing, among others, 
file metadata (e.g., hashes, size),
certificate metadata for signed samples
(e.g., thumbprint, subject),
VT specific data
(e.g., time of first submission to VT, submission filenames), and
the list of detection labels assigned by up to 70 antivirus (AV) engines
used to scan the file.

AV labels in VT reports are routinely used by security analysts 
for determining if a file is malicious by applying a threshold on the 
number of detections~\cite{zhu2020measuring} and for 
identifying the family of a sample~\cite{avclass}. 
Similarly, malware developers often leverage VT during 
development to check if their samples are detected 
and, if so, revise them until they 
become \emph{fully undetected} 
(FUD)~\cite{graziano2015needles,huang2016android,yuan2019towards}.
VirusTotal is also commonly used as a source for collecting malware 
samples~\cite{huang2016android,cozzi2018understanding,mantovani2020prevalence,maffia2021longitudinal,alrawi2021circle}. 
However, downloading specific samples from VT requires a privileged API key 
with a strict limit in the number of downloads and 
there exist other malware repositories with more lax download 
requirements~\cite{virusshare,malshare,malwareBazaar}.
 
Amongst its more recent commercial services, 
VT offers a file feed 
(hereafter, the VT file feed or simply the feed),
a stream of analysis reports for all submissions~\cite{vtfeedapi}.
The feed service also allows unlimited downloads of the samples 
submitted in the last seven days. 
Thus, the VT file feed could be used as a source of malware for building 
malware datasets such as those required by 
machine learning (ML) based 
malware detection (e.g.,~\cite{drebin,smutz2012malicious,mtnet,malconv,neurlux}) and 
family classification (e.g.,~\cite{rieck2008learning,mtnet}),
as well as for 
malware ecosystem studies (e.g.,~\cite{cozzi2018understanding}).
However, the VT file feed has never been characterized in depth so 
there remain key questions about it 
such as
How diverse is it? 
Does it allow building malware datasets for different filetypes? 
How fresh are the samples it provides? 
What is the distribution of malware families it sees? 
Does that distribution really represent the distribution of malware on 
user devices?

To address these questions, we analyze one year of VT file feed reports, 
from \observationStart to \observationEnd. 
During the first 11 months we collect reports where the sample is detected 
by at least one AV engine, while in the last month we collect all 
feed reports, regardless of the number of detections.
Our dataset comprises of 328M reports for 235M samples, 
of which 209M samples are new (i.e., previously unknown) to VT.
To analyze how well the VT file feed represents the 
malware ecosystem, we intersect the feed data with 
telemetry data collected in a privacy-sensitive manner from 
tens of millions of devices of clients of a large \vendor.

VT is also a great dataset for hunting new threats, 
as demonstrated by previous works that detect malware 
developments~\cite{graziano2015needles,huang2016android,yuan2019towards} and
perform malware triage~\cite{spotlight}.
However, the stream of VT submissions is massive, 
which raises questions about how to 
scale threat hunting to the feed contents 
without applying extremely aggressive filtering.
This question equally applies to other massive file feeds such as those 
 collected by \vendors~\cite{ugarte2019close}.
To address this issue, this paper investigates 
scalable approaches for clustering the 235M feed samples.
We posit that large-scale threat hunting requires 
highly scalable clustering approaches that produce high precision 
(i.e., high purity) clusters.
High precision clusters contain a vast majority of related samples 
(i.e., samples belonging to the same family).
They allow the analyst to examine only a few samples in 
the cluster and generalize the analysis results to the whole cluster.
This property holds even if the clusters have low recall
(i.e., a family is split among multiple clusters).
On the other hand, low precision clusters are problematic for generalizing 
from a few to all samples in the cluster.
We investigate three malware clustering approaches:
hierarchical agglomerative clustering (HAC), 
a more scalable HAC variant for TLSH digests (HAC-T)
that reduces the number of comparisons from quadratic to 
$\mathcal{O}(n\log{}n)$~\cite{oliver2020hac}, and 
a simple feature value grouping (FVG) that groups samples 
on the value of one feature,
e.g., all samples with the same import hash~\cite{imphash} form a cluster.
Our results show that HAC-T and FVG using selected features 
produce high precision clusters (0.90--0.99) on ground truth datasets. 
However, only FVG scales to the daily influx of samples in the feed. 
Moreover, FVG takes 15 hours to cluster the whole dataset of 235M samples. 
We use the produced clusters for threat hunting, 
namely for detecting 190K samples thought to be benign 
(i.e., with zero detections) that
may really be malicious because they belong to 29K clusters where
most samples are detected as malicious.

The following are the most significant contributions:

\begin{itemize}

\item We characterize the VT file feed using 328M analysis reports 
for 235M samples collected over one year, 
as well as the telemetry of a \vendor on the same time period.
Among others, we show that despite having a volume 17 times 
lower than the telemetry, 
the feed observes 8 times more malware.
The feed is fresh; it receives over one million new samples each day and 
samples appear a median of 4.4 hours after they are 
seen in user devices.
Of the new samples, 62\% are variants of known families that can be labeled 
on first sight, but 38\% are unlabeled.
0.3\% of new samples are originally FUD, 
i.e., they have no detections on first sight, but later multiple AV engines 
start flagging them.
The feed is diverse containing a wealth of filetypes and 
4.9K families with at least 100 samples. 
However, the diversity is largely due to Windows and Android families.
On the negative side, the family distribution 
significantly differs from that observed in the telemetry indicating that
VT may not be a good source to infer family impact on real users.

\item We explore clustering approaches that can scale to the 
hundreds of millions of samples in the VT file feed. 
We evaluate three clustering approaches: 
hierarchical agglomerative clustering (HAC), 
a more efficient HAC version for the TLSH digest (HAC-T), and 
a simple grouping by feature value (FVG).
We first evaluate the accuracy using Windows and Android ground truth datasets.
Then, we evaluate scalability showing that   
only FVG scales to the feed volume.

\item We apply FVG clustering to the problem of detecting 
fully undetected malware.
Using FVG, we cluster 235M samples regardless if they are 
detectected by AV engines. 
We identify 29K FVG-vhash clusters with a majority of 
samples detected by multiple AV engines, 
containing 190K likely malicious samples with zero detections.

\end{itemize}
\section{Data Collection}
\label{sec:collection}

The VT file feed contains all analysis reports for samples submitted to VT. 
These include reports for new file samples (i.e., first submission to VT),
resubmissions of previously submitted samples, and 
user-requested re-scans of previously submitted samples.
Thus, multiple reports may appear in the feed for the same sample. 
Throughout the paper, we use \emph{new samples} to denote samples 
that were first submitted to VT during our collection period
(the first submission date of a sample is included in its reports).
In general, we focus on the last report we collected for each sample 
because it should provide the most up-to-date information
(e.g., updated AV labels). 
However, when we are interested in what happened to a sample when first 
submitted to VT (e.g., whether it was detected or labeled) 
we will focus on the first report of new samples.

\begin{table}[t]
\small
\centering
\begin{tabular}{l|r|rrr}
\hline
\textbf{Data}  & \textbf{All} & \textbf{peexe} & \textbf{apk} & \textbf{other}\\
\hline
Reports			&  328.3M  &  220.3M & 15.9M &  92.0M \\
Samples			&  235.7M  &  155.5M &  8.2M &  72.0M \\ 
New Samples		&  209.6M  &  134.6M &  5.6M &  69.3M \\
\hline
\end{tabular}
\caption{Summary of dataset collected from the VT file feed 
between \ostart and \oend.
}
\label{tab:datasets}
\end{table}

\ignore{
\begin{table*}[t]
\small
\centering
\begin{tabular}{lrrrr}
\hline
\textbf{Property} & \textbf{All} & \textbf{D1} & \textbf{D2} & \textbf{D3} \\
\hline
Start Date    & 2020/12/21  & 2020/12/21 & 2021/02/08  & 2021/05/21\\
End Date      & \review{2021/07/01}  & 2021/01/10 & 2021/05/15  & \review{2021/07/01} \\
Days          & 158        		& 21         &    96      & 41 \\
Reports       &  				& 22,972,194 &  & \\
Samples       & 				& 19,047,982 &   & \\
\, Signed     & 5,331,923   	& 1,224,560  &  3,009,841 & 1,097,522 \\
Certificates  & 775,125			& 177,086    &  366,719   & 231,320 \\
\hline
\end{tabular}
\caption{Summary of VT feed dataset.
\kevin{My results for D1 are: Reports: 23,043,611 Samples: 18,706,011. 
For reports, I have counted the number of AvClass2 daily files lines 
during the selected period. 
For Samples, I have created a set of the SHA2 hash from every line.}
}
\label{tab:datasets}
\end{table*}
}

We collect reports from the feed every minute. 
Due to storage constraints  
we do not download the samples from the feed, only the reports.
Currently, we store the complete VT reports on disk in compressed form.
However, this may not be sustainable on the long run.
An alternative would be to store only the parts of the VT report needed to
generate our feature files, described in Section~\ref{sec:features}.
To keep the storage manageable, 
in the first 11 months we were only collecting reports 
where at least one AV engine detected the file 
as malicious, which roughly corresponds to half of all reports in the feed.
On November 19th, 2021, we scaled up our collection infrastructure 
and started collecting all reports in the feed 
regardless of the number of detections, 
i.e., including reports with zero detections. 
Overall, as summarized in Table~\ref{tab:datasets},
over one year between \observationStart and \observationEnd, 
we collected 328M reports for 235M samples (by unique file SHA256), 
of which 209.6M (89\%) samples were new to VT, 
i.e., first seen by VT during our collection.

\paragraph{Telemetry.}
In Section~\ref{sec:vtfeed}, 
we compare the VT file feed volume and collection delay to the 
telemetry of a large \vendor collected over the same one year period.
The telemetry comes from tens of millions of real 
Windows devices in use by customers of the {\vendor}'s AV engine. 
The AV engine queries a central service with file hashes observed 
on the device to obtain file reputation information.
The customers opted-in to sharing their data and the devices are anonymized
to preserve customer privacy.
The telemetry only contains file metadata, it does not contain the samples.

\begin{table}[t]
\small
\centering
\begin{tabular}{lllcc}
\hline
\textbf{Feature} & \textbf{Scope} & \textbf{Type} & \textbf{peexe} & \textbf{apk} \\
\hline
authentihash & sample & cryptohash & \Y & \N \\ cert\_issuer & sample & string & \Y & \Y \\ cert\_subject & sample & string & \Y & \Y \\ cert\_thumbprint & sample & cryptohash & \Y & \Y \\ cert\_valid\_from & sample & timestamp & \Y & \Y \\ cert\_valid\_to & sample & timestamp & \Y & \Y \\ exiftool\_filetype & sample & string & \Y & \Y \\ fseen\_date & sample & timestamp & \Y & \Y \\ icon\_hash & sample & cryptohash & \Y & \Y \\ imphash & sample & cryptohash & \Y & \N \\ md5 & sample & cryptohash & \Y & \Y \\ package\_name & sample & string & \N & \Y \\ richpe\_hash & sample & cryptohash & \Y & \N \\ sha1 & sample & cryptohash & \Y & \Y \\ sha256 & sample & cryptohash & \Y & \Y \\ tlsh & sample & fuzzyhash & \Y & \Y \\ trid\_filetype & sample & string & \Y & \Y \\ vhash & sample & structhash & \Y & \Y \\ \hline
detection\_labels & scan & string list & \Y & \Y \\ scan\_date & scan & timestamp & \Y & \Y \\ sig\_verification\_res & scan & string & \Y & \N \\ vt\_meaningful\_name & scan & string & \Y & \Y \\ vt\_score & scan & integer & \Y & \Y \\ \hline
avc2\_family	& derived & string & \Y & \Y \\ avc2\_tags		& derived & string list & \Y & \Y \\ avc2\_is\_pup	& derived & bool & \Y & \Y \\ filetype & derived & string & \Y & \Y \\ \hline
\end{tabular}
\caption{Features used.}
\label{tab:features}
\end{table}

\ignore{
\begin{table}[t]
\small
\centering
\begin{tabular}{lp{6cm}}
\hline
\textbf{Feature} & \textbf{Input fields} \\
\hline
filetype & trid\_file\_type, exiftool\_filetype, sig\_verification\_res \\ 
family & detection\_labels \\ 
program name & exiftool\_original\_fname, authcode\_orig\_name, vt\_meaningful\_name \\ 
\hline
\end{tabular}
\caption{Features derived for each VT report.}
\label{tab:derived}
\end{table}
}

\section{Features}
\label{sec:features}

Since we do not download the samples,
we need to restrict our analysis to features available in the reports, 
or that can be derived from the reports.
We focus on a selected set of \numfeatures: 
\nvtfeatures from the VT reports
and \nderivedfeatures derived from those
(e.g., filetype and malware family).
Features are summarized in Table~\ref{tab:features}. 
We define three scopes for a feature: sample, scan, and derived. 
Sample features should have the same value across all scans of a sample. 
On the other hand, scan features 
may differ across scans of the same sample, 
i.e., they evolve over time.
For example, the hash of the certificate of a signed sample 
(\feature{cert\_thumbprint}) should always be the same. 
But, whether the signature of a signed sample validates 
(\feature{sig\_verification\_res})
can change across scans, 
e.g., if the certificate expires or is revoked.
Features may be extracted only for a subset of filetypes, 
e.g., be specific to Windows PE executables or Android APKs, and 
may be null for some samples
(e.g., certificate features are not available for unsigned Windows executables).
We detail the VT report features in Section~\ref{sec:vtFeatures} and 
the derived features in Section~\ref{sec:derivedFeatures}. 

\subsection{VT Report Features}
\label{sec:vtFeatures}

\paragraph{Hashes.}
Our features include three types of hashes: 
cryptographic, fuzzy, and structural.
Cryptographic hashes include three hashes applied over the whole file
(\feature{sha256}, \feature{sha1}, \feature{md5})
and five hashes calculated over parts of a file:
\feature{imphash} covers the import table of a 
PE executable~\cite{imphash};
\feature{richpe\_hash} covers the optional Rich Header 
that contains information about the compilation chain
of PE executables~\cite{webster2017finding};
\feature{cert\_thumbprint} covers the leaf certificate for signed samples;
\feature{authentihash} covers a PE executable excluding 
Authenticode code signing fields; and
\feature{icon\_hash} covers the icon image optionally 
embedded in Windows and Android files.
All cryptographic hashes are compared using equality, 
i.e., two files have the same hash or not.

VT reports contain two fuzzy hashes calculated over the whole file:
\feature{tlsh} and \feature{ssdeep}. 
We favor \feature{tlsh} because prior work has shown that \feature{tlsh} 
generally outperforms \feature{ssdeep}~\cite{pagani2018beyond} and 
because a scalable clustering for TLSH digests has been recently proposed~\cite{oliver2020hac}.
\feature{tlsh} uses Hamming distance on the 
hash value to determine if two inputs are similar.

Our features also include the VirusTotal hash (\feature{vhash}), 
VT's proprietary structural hash.
According to VT's minimal documentation on it~\cite{vhash},
it takes into account sample properties such as
imports, exports, sections, and file size. 
From our evaluation, \feature{vhash} appears to be a common name for a set of
structural hashes for different filetypes.
VT computes \feature{vhash} for many filetypes
(e.g., PE executables, APKs, PDFs, Office documents), 
but not for all (e.g., JPEG and PNG images have no \feature{vhash}). 
VT provides daily clusters of files using \feature{vhash}. 
We have checked that samples in the same cluster have the 
same \feature{vhash} value, 
so we use equality to compare two \feature{vhash} values.
While proprietary and largely undocumented, our experimental evaluation 
in Section~\ref{sec:clustering} shows that \feature{vhash} is a 
good feature for clustering.

\ignore{
\feature{tlsh}~\cite{tlsh} was proposed by TrendMicro.
It slides a windows of size 5 bytes over the input string
to populate an array of bucket counts.
Then, it calculates the quartile points.
The distance function between two digests is 
the sum of the distance of the headers
and the distance of the digest bodies,
the latter being a variation of the Hamming distance.
We normalize the distance into a [0-1] range adopting the approach
in~\cite{upchurch2015variant}.
\feature{ssdeep}~\cite{ssdeep} uses Context Triggered Piecewise Hashing
(CTPH) to segment a file producing a 6 bit hash value for each segment.
Digests are compared using edit distance,
which we normalize into a [0-1] range.
\feature{dhash}~\cite{dhash} (difference hash) is a perceptual hash 
that produces similar hash values for images that look 
similar to a human (e.g., resized, color changes).
Digests are compared using Hamming distance,
which we normalize into a [0-1] range by dividing it by the
maximum length of two hashes in bits.
}

\paragraph{Timestamps.}
We obtain four timestamps from a VT report. 
The \feature{scan\_date} when the sample is analyzed,
which is always within our collection period.
The VT first seen date (\feature{fseen\_date})
when the sample was first submitted to VT.
For signed samples, we also obtain  
the certificate's validity period defined by the 
\feature{cert\_valid\_from} and \feature{cert\_valid\_to} dates. 

\paragraph{AV scans.}
VT scans each submitted sample with a large number of AV engines. 
We extract the number of engines that detected the sample 
(i.e., gave it a non-NULL label) (\feature{vt\_score})
and the list of \feature{detection\_labels}. 
The labels are used to derive three classification features, 
as detailed in Section~\ref{sec:derivedFeatures}. 

\paragraph{Program names.}
We use two features that capture the program a sample corresponds to.
The \feature{package\_name} is the package identifier for Android apps and 
\feature{vt\_meaningful\_name} is the most meaningful filename VT selects 
for a sample
(e.g., among all filenames of the sample when submitted to VT).

\subsection{Derived Features}
\label{sec:derivedFeatures}

\paragraph{Filetype.}
Determining the filetype of the sample in a report is not straightforward
because VT reports do not have a single field for it. 
Instead, there are multiple fields that provide, 
possibly contradictory, filetype information.
We derive a unique \feature{filetype} feature for each report by performing a 
majority voting on three fields: 
\feature{trid\_file\_type}, 
\feature{vt\_tags}, and 
\feature{vt\_meaningful\_name}.
\feature{trid\_file\_type} captures the filetype identified by the 
\trid tool~\cite{trid}, 
which has very fine-grained granularity 
(e.g., over 90 \emph{peexe} subtypes).
We build a mapping from \trid filetypes to coarser-grained filetypes 
such as grouping all Windows PE files 
(e.g., EXE, DLL, OCX, CPL) under \emph{peexe} and 
all Word files (DOC, DOCX) under \emph{doc}.
\feature{vt\_tags} provides a list of tags assigned by VT 
to enable searching for samples across different dimensions.
Some of the tags such as \emph{apk}, \emph{peexe}, and \emph{elf} provide 
filetype information.
When \feature{vt\_meaningful\_name} is available, 
we extract the extension from the filename and map the extension to a filetype. 

\paragraph{AVClass features.}
We feed the \feature{detection\_labels} to the \avclass malware 
labeling tool~\cite{avclass2}.
\avclass outputs 
a list of tags (\feature{avc2\_tags}) for the sample that include its 
category, behaviors, file properties, and 
the most likely family (\feature{avc2\_family}). 
It also provides whether the sample is considered 
potentially unwanted or malware (\feature{avc2\_is\_pup}).
\avclass uses a taxonomy to identify non-family tokens that may appear 
in the AV labels such as 
malware classes (e.g., \emph{CLASS:virus}),
behaviors (e.g., \emph{BEH:ddos}), 
file properties (e.g., \emph{FILE:packed:asprotect}), and 
generic tokens (e.g., \emph{GEN:malicious}). 
It also uses tagging rules to identify aliases between families 
(e.g., \emph{zeus} being an alias to \emph{zbot}).
In this work we apply \avclass to 328M VT reports, 
eight times more than the largest to date work~\cite{avclass2}. 
Thus, our \avclass results include a wealth of new tags, 
including new aliases and non-family tokens. 
We have used the \avclass update module and extensive manual validation 
to identify new tagging rules that 
capture previously unknown aliases, as well as new taxonomy entries for 
tokens appearing in over 100 samples. 
This process has resulted in doubling the \avclass taxonomy from 
1,354 entries in the \avclass repository to 2,828 entries in our 
version and a 58\% increase in tagging rules 
from 1,579 to 2,496 entries.

\section{Feed Analysis}
\label{sec:vtfeed}

This section characterizes the VT file feed, 
answering the following questions:
(1) How large is the VT file feed in comparison to the AV telemetry?
(2) How fresh are samples in the feed? (3) How diverse is the feed in terms of filetypes? (4) What fraction of samples can be detected as malicious on first scan?
(5) How diverse is the feed in terms of families? (6) What fraction of malicious samples are fully undetected on first scan?
(7) What fraction of samples are signed?
(8) What is the overlap in malicious samples between the feed and 
the AV telemetry? 
And which one sees samples faster?

\begin{figure}[t]
\centering
\includegraphics[width=\columnwidth]{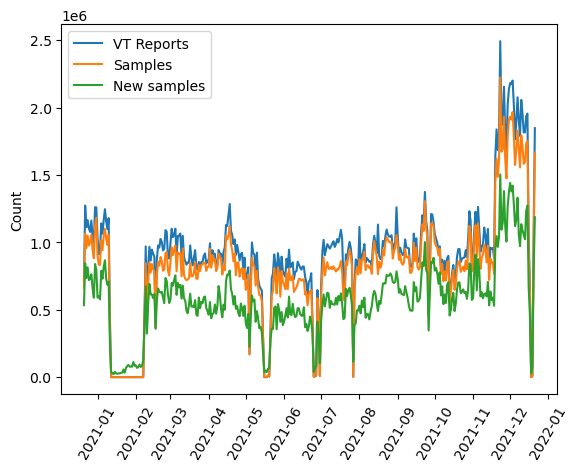}
\caption{Number of daily VT reports and samples collected.}
\label{fig:daily_vt_reps}
\end{figure}

\paragraph{Volume.}
Figure~\ref{fig:daily_vt_reps} shows for each day in the collection period,
the number of reports in the feed,
the number of unique samples in the daily reports, and
the number of samples first seen by VT on that day.
The figure shows a few gaps when the collection infrastructure was not working,
the longest taking place between January 11th and February 7th.
The volume of reports and hashes significantly increases 
once we started collecting samples with no detections.
We compute the daily statistics, 
excluding days in the collection gaps, 
split into two periods: 
before \fullCollectionStart when we were collecting 
only reports with at least one detection, and 
after that date when we were collecting all reports.
Table~\ref{tab:stats_daily_after} shows the daily stats 
when collecting all reports: 
the average number of daily reports is nearly 1.7M,
the average number of samples nearly 1.5M, and
the average number of new samples 1.0M.
When only collecting reports with at least one detection
the averages were
811K reports, 732K samples, and 515K new samples.
Thus, approximately half of the 
reports (48\%), 
samples (49\%), and 
new samples (50\%) in the feed are benign.
 
\begin{table}[t]
	\small
	\centering
	\begin{tabular}{l|r|rrr}
		\hline
		& \textbf{Mean} & \textbf{Median}	& \textbf{Stdev}	&	\textbf{Max}\\
		\hline
		Reports			&  1,681,470	& 1,879,952 & 632,366	& 2,492,454	\\
		Samples	    &  1,493,410	& 1,680,520 & 558,895	& 2,223,638	\\
		New samples	&  1,028,370	& 1,120,242	& 382,196	& 1,504,174	\\
		\hline
	\end{tabular}
	\caption{Daily statistics when collecting all feed reports 
  (from \ochange to \oend).
}
	\label{tab:stats_daily_after}
\end{table}

We compare the feed volume when collecting all reports 
with the telemetry data over the same period.
Over that month, 
the VT file feed contains reports for 39.8M samples,
while the telemetry contains queries for 686.5M samples. 
Both numbers include all samples observed in the time period, 
regardless if the samples are old or new and 
whether they are benign or malicious. 
Thus, the telemetry volume is 17 times larger than the VT file feed volume.
This is a lower bound with respect to the whole file volume 
available to the \vendor, as in addition to the Windows endpoint telemetry, 
the \vendor has also other datasets.
On the other hand, the telemetry captures that over that month, 
the AV engine threw alerts for 1.9M malicious files in 905K devices.
In comparison, the VT file feed contains 14.8M samples
with at least four detections on the same time period, 
nearly eight times more detected malware.
This discrepancy is likely due to two reasons. 
First, prior work has shown that 
telemetry data is largely dominated by rare benign files, 
i.e., 94\% of files in AV telemetry 
are observed only in one device and the ratio of benign
to malicious such files is 80:1~\cite{li2017large}.
Second, if we assume that the telemetry represents the file distribution 
in the contributing devices, 
then the VT file feed is clearly biased towards malicious samples. 
This makes sense as VT is widely considered a threat sharing platform. 
Thus, contributors may share only suspicious samples, 
more likely to be malicious. 
Furthermore, contributors may purposefully avoid known benign samples since 
they are more likely to contain private data.

\take{At the end of 2021, the VT file feed had daily averages of 
1.7M reports, 1.5M samples, and 1.0M new samples. 
While massive, the feed volume is 17 times lower than 
the telemetry of a \vendor. 
However, despite the much lower volume, 
the VT file feed contains eight times more malware than the telemetry, 
making it a great source of malicious samples.
}

\paragraph{New samples.}
Table~\ref{tab:datasets} shows that 89\% of the observed samples are new,
i.e., are first observed by VT after we started collecting.
By filetype, 61\% \emph{peexe} samples are new, 
compared to 35\% for \emph{apk} samples, and 
81\% for other samples.
This indicates lower polymorphism for APKs 
(e.g., less versions for benign programs and/or less obfuscation for malware)
making it easier for multiple users to submit the same sample to VT.

\begin{figure}[t]
  \centering
  \includegraphics[width=\columnwidth]{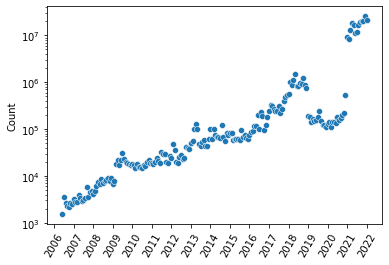}
  \caption{Number of samples first seen by VT on each month.
  y-axis is in logarithmic scale.
    }
  \label{fig:fseen_vt_month}
\end{figure}

The VT first seen date provides a lower bound for a sample's lifetime, 
i.e., the sample could be older if it took some time for it to be submitted 
to VT.
The oldest sample observed in our collection period was first seen by 
VT on May 22nd, 2006. 
Figure~\ref{fig:fseen_vt_month} shows the number of samples 
(in logarithmic scale) whose VT first seen date is on each month,
capturing how old are the samples already known to VT.
The shape of the figure captures the volume increase in samples submitted 
to VT over time until 2019, followed by a decrease in 2019-2021. 
The reduction 
could be due to some vendors reducing their sharing from 2019.

\take{Of the collected samples 89\% are new, i.e., previously unknown to VT, 
and the feed provides over a million new samples each day.
Thus, the VT file feed is a great source of fresh samples. 
}

\begin{table}[t]
	\small
	\centering
	\begin{tabular}{lrr}
		\hline
		\textbf{Filetype} & \textbf{Samples} & \textbf{Perc} \\
		\hline
peexe   &       155,526,594 &     65.97\% \\
javascript      &       21,048,404 &      8.93\% \\
html    &       12,540,571 &      5.32\% \\
pdf     &       11,346,815 &      4.81\% \\
apk     &       7,992,206 &       3.40\% \\
text    &       5,149,050 &       2.18\% \\
NULL    &       4,128,183 &       1.75\% \\
zip     &       3,934,987 &       1.67\% \\
dex     &       3,015,650 &       1.28\% \\
gzip    &       2,926,739 &       1.24\% \\
lnk     &       2,718,635 &       1.15\% \\
elf     &       942,148 &        0.40\% \\
rar     &       516,514 &        0.22\% \\
jar     &       448,324 &        0.19\% \\
doc     &       429,794 &        0.18\% \\
xls     &       428,057 &        0.18\% \\
macho   &       409,399 &        0.17\% \\
php     &       352,143 &        0.15\% \\
xml     &       335,962 &        0.14\% \\
powershell      &       321,178 &        0.14\% \\
		\hline
		Other 		& 1,233,754    	& 0.52 \% \\
		\hline
		ALL 		& 235,745,107	& 100.0\% \\
		\hline
	\end{tabular}
	\caption{Top 20 filetypes for all samples observed.
		\emph{peexe} includes all Windows PE files (EXE, DLL, CPL, OCX, ...)
		\emph{doc} and \emph{xls} include also \emph{docx} and \emph{xlsx}, 
		respectively.
		\emph{NULL} corresponds to samples for which a filetype could not be determined.
			}
	\label{tab:ftypes-samples}
\end{table}

\paragraph{Filetypes.}
Table~\ref{tab:ftypes-samples} shows the top 20 filetypes for all samples 
observed.
The feed is dominated by Windows PE files (EXE, DLL, OCX, CPL, ...) 
that correspond to 66\% of the samples.
Far behind are other filetypes like 
JavaScript (8.9\%),
HTML (5.3\%),
PDF (4.8\%), and
Android applications (3.4\%).
These top 5 filetypes cover 88.4\% of all samples.
We could not obtain a filetype for 1.7\% of samples 
as they had no \trid information, no VT filetype-related tags, 
were not signed, and had no most meaningful filename with extension.
This highlights the lack of a unified filetype field and the 
limitation of the tools VT uses for filetype determination. 

Ugarte-Pedrero et al.~\cite{ugarte2019close} reported that 51\% of 
an AV feed were PE executables. 
The larger VT file feed ratio may be due to users contributing more frequently 
PE executables to VT, avoiding other filetypes like HTML or text files 
that may contain more private data.

\take{Two thirds of feed samples are Windows PE files, 
but the feed is a good source of samples for a large variety of filetypes.
The feed lacks a unified filetype field and filetype identification is 
challenging for a significant number of samples.}

\begin{figure}[t]
	\includegraphics[scale=.5]{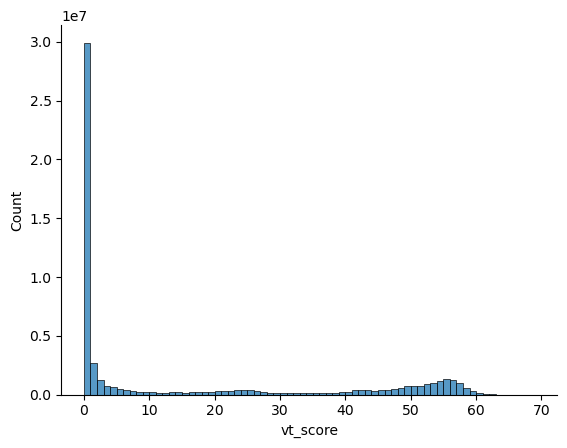}
	\caption{Number of detections distribution for all reports since 2021/11/19.}
	\label{fig:vtc_distribution_since_change}
\end{figure}

\begin{figure}[t]
	\includegraphics[scale=.5]{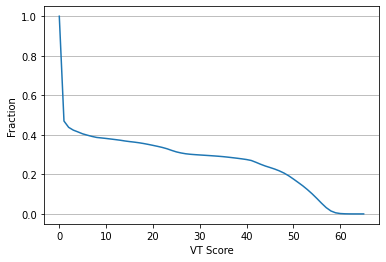}
	\caption{Reverse ECDF for the first report of each new sample since 2021/11/19.}
	\label{fig:reverse_ecdf}
\end{figure}

\paragraph{AV detections.}
A common approach for detecting malicious samples is to apply a threshold on
the number of detections in a VT report~\cite{zhu2020measuring}.
We use this approach to quantify the percentage of 
malicious samples in the feed.
We focus on the last month when collecting all feed reports.
Figure~\ref{fig:vtc_distribution_since_change} 
shows the distribution of the number of 
AV detections for all reports collected starting \ochange.
The figure shows that 85\% of the reports in the last month have no detections 
and 7\% have one detection. 
But, there are 9.6M samples with at least 40 detections. 

We also examine the number of detections the first time a sample is 
submitted to VT. 
Figure~\ref{fig:reverse_ecdf} shows the complementary CDF of VT scores 
for the first report of each new sample since \ochange. 
The figure captures the fraction of malicious samples in the feed depending on 
the selected detection threshold. 
53\% of the samples have zero detections on their first observation. 
This percentage includes truly benign programs as well as 
malicious samples that go fully undetected. 
If we set the detection threshold on at least one detection, 
47\% of the samples would be considered malicious. 
If the threshold is set higher to minimize false positives, 
that reduces the fraction of malicious samples, 
e.g., 41\% if we set it to at least four detections 
as done in several related works~\cite{malsign,ppipup,andropup}. 

\take{The VT file feed is a file feed rather than a malware feed. 
Half of its volume is for undetected samples and 
53\% of the samples have no detections on first observation. 
On the other hand, 41\%--47\% of samples can be detected as malware 
by the AV engines on first sight because they 
share traits with previously seen malware 
(i.e., match existing signatures).
}

\paragraph{Labeling.}
We obtain a sample's family by feeding to \avclass 
the last report of each sample in our dataset,
which should have the most up-to-date labels.
\avclass labels 151.7M (64.3\%) of the samples 
with 74,360 distinct family names.
However, many families output by \avclass are rare. 
In particular, 41.4K (56\%) of all families have only one sample, 
14K (19\%) have at least 10 samples,
4.9K (7\%) have at least 100 samples, and 
only 1.5K (2\%) have at least 1,000 samples.
Despite more than half of the families having only one sample, 
the fact that there are 4.9K families with more than 100 samples 
shows that the feed is diverse and 
is not dominated by a few highly polymorphic families (e.g., file infectors).
However, the diversity is largely due to Windows families. 
By filetype, the number of families with more than 100 samples is 
led by \emph{peexe} with 3.8K families, followed far behind by
\emph{apk} (447), 
\emph{html} (129), 
\emph{javascript} (116),
\emph{doc} (53),
\emph{macho} (52), 
\emph{xls} (47),
\emph{elf} (37), and 
\emph{pdf} (15).
Thus, by monitoring the feed it is possible to build datasets 
with a large number of families for Windows and Android malware. 
But, for other filetypes like \emph{macho} and \emph{elf}, 
even after collecting for a year, 
we could only obtain 52 and 37 families with at least 
100 samples, respectively.

\avclass outputs as family 
the top-ranked tag that is either a family in the taxonomy or unknown
(i.e., not in the taxonomy).
Of the 74,360 families output by \avclass, 
2,391 (3.2\%) are in the updated taxonomy,
which contains a total of 2,451 families
(i.e., 97.5\% of taxonomy families appear in one year of feed reports).
However, the families in the updated taxonomy contribute 90.6\% of the labeled 
samples, only 9.4\% of the samples are labeled with unknown families.
This indicates that the most popular families are in the updated taxonomy, 
which is expected as it is common for analysts like us to add the most 
popular previously unknown families to the taxonomy. 
In fact, of the families with at least 1M samples, only 3\% are unknown, 
increasing to 
15\% for families with 100K samples,  
43\% for those with 1K samples, and
85\% for those with 10 samples.
Unknown families can be due to two main reasons.
One are tags that it is unclear if they are a family name 
or another category such as a behavior or a file property
(e.g., \emph{lnkrun}, \emph{refresh}).
The other are tags that correspond to random-looking signature identifiers or 
family variants
(e.g., \emph{aapw}, \emph{dqan}).
We manually examine the top 1K families and identify 
that 89\% of the unknown families correspond to the first case and 11\% to the latter.
We repeat this check on 200 randomly sampled unknown families with only 
one sample and the result is the opposite: 
11\% corresponding to the first case and 89\% to the latter.
Thus, for less prevalent families \avclass may output a name that 
corresponds to a signature identifier or variant. 
While those random-looking names are not very descriptive for analysts, 
they are still valid cluster identifiers, 
i.e., samples with the same name should belong to the same family.
Based on the above, we estimate that over the whole year a total of 33.8K 
(41.4K * 0.11 + 32.9K * 0.89)
families of all filetypes have been observed in the feed.

We also obtain the family using the first report for new samples. 
\avclass is able to label on first sight 62.3\% of 
new samples, slightly less than the 64.3\% using the last collected report.
It is worth noting that the percentage of malicious samples labeled 
tends to increase over time as illustrated in 
Section~\ref{sec:clustering} where \avclass labels 97.8\% of the 
samples in the Malicia dataset from 2012~\cite{malicia} and 98.9\% of the 
samples in the AMD dataset from 2016~\cite{amd}.
The fact that 62\% of new samples can be attributed on first sight to a
family indicates they they correspond to variants of known families 
with accurate signatures.
This result shows that \avclass can be used as a filter to remove 62\% 
of samples from well-detected families so that analysts 
can focus on the 38\% unlabeled samples.

Prior work has applied \avclass to \emph{peexe}, \emph{apk}, and 
\emph{elf} files (e.g.,~\cite{avclass2,cozzi2018understanding}).
However, \avclass can be applied on AV labels regardless of 
platform or filetype.
Table~\ref{tab:families_per_filetype} 
shows the top 10 families for the six executable and macro filetypes.
The largest families overall are for Windows led by \emph{berbew} 
with 19.4M samples, followed by \emph{dinwod} (9.4M), and 
\emph{virlock} (7.9M).
We use \avclass to output a relations file on the whole feed.
We identify a family's class checking the strongest CLASS relation 
for each family with a strength of at least 0.2.
The top 10 \emph{peexe} families are dominated by 4 worm and 3 virus families
due to their high polymorphism.
However, as already discussed, 
overall the feed is not dominated  by file infectors and worms. 
For Android, the top 10 families are all PUP and 8 of them are adware.
The top Linux families are dominated by backdoors 
including \emph{mirai} derivatives 
(\emph{gafgyt}, \emph{hajime}, \emph{mozi}).
For macOS, seven top families are PUP and five of those adware. 
Finally, Word and Excel macros are dominated by downloaders. 
Table~\ref{tab:other_families_per_filetype} 
shows the top families for three other popular filetypes 
(JavaScript, HTML, PDF) 
for which we observe that top families output by \avclass 
contain many unknown tokens that may correspond instead to other categories 
(e.g., \emph{redir} may indicate injections that redirect the user). 
We also observe overlaps between JavaScript and HTML families 
(e.g., \emph{cryxos}, \emph{facelike}) and that for 9/30 families 
\avclass cannot identify a class.
We conclude that for these three filetypes the concept of a family 
is not as well defined and that AV labels for these filetypes focus instead on  
concepts such as phishing, injections, and exploitation. 

\take{The feed is diverse with 4.9K families with at least 100 samples.
However, the diversity is largely due to \emph{peexe} and \emph{apk} families.
For those two filetypes, the feed is a good source to build datasets 
for large-scale family classification. 
\avclass labels 62\% of samples on first sight. 
Thus, it can be used as a filter to remove samples
from well-detected families so that analysts can focus on the 
38\% unlabeled samples.
}

\paragraph{Originally FUD malware.}
It is possible that a malicious sample is fully undetected
when first submitted to VT, but a later report classifies it as malicious.
We detect such \emph{originally FUD} samples here and 
address the detection of FUD malware 
with zero detections to date in Section~\ref{sec:hunting}.
To detect originally FUD samples,
we measure the number of samples that satisfy three conditions: 
(1) they are new samples first observed by VT during our collection period;
(2) their last report has at least 4 detections; and 
(3) their first report had zero detections \emph{or}
their VT first seen date is not in a data collection gap and 
is before their first observation. 
The last condition is a disjunction to address that  
we only collected reports with zero detections in the last month. 
During the first 11 months we can know if a sample 
had zero detections in their first scan because their VT first seen date 
is in our collection period and happens before the earliest scan date 
collected for the sample.
The exception are samples first seen during a collection gap, 
for which a delayed scan date does not necessarily imply 
zero detections on the first scan.

We identify 637K samples satisfying those conditions. 
However, the time difference between the 
first seen date and the first report with at least four detections, 
indicates that 37K samples change from zero to at least four detections 
within 5 minutes of their first VT observation. 
We exclude those 37K samples as we observe that  
the distribution stabilizes afterwards 
(i.e., after 15 minutes only an extra 1K samples flip classification).

Thus, we identify 600K originally FUD samples
(0.3\% of all new samples)
that had no detections on their first scan, 
but were later considered malicious by at least 4 AV engines. 
Increasing the detection threshold would decrease the percentage, 
but the detection rate of a malicious sample tends to increase over time and 
for 82\% of samples we only have one report.
Thus, we believe our FUD rate estimation is conservative.
The median time to flip classification is 7 days, 
(mean of 23.8 days)
with 12\% of the samples flipping classification in less than one day.

Of the 600K originally FUD samples, 
60\% are \emph{peexe}, followed by 
11\% \emph{pdf}, and
8\% \emph{javascript}.
PDFs are more than twice as likely to be FUD than expected  
since they comprise only 4.8\% of all feed samples. 
Malicious PDFs typically contain exploits and are 
used in spearphishing attacks.
These numbers point to malicious PDFs being harder to detect.

\begin{table}[t]
\small
\centering
\begin{tabular}{lllrr}
\cline{4-5}
\multicolumn{3}{c}{} & \multicolumn{2}{c}{\textbf{FUD}} \\
\hline
\textbf{Family} & \textbf{Class} & \textbf{Filetype} & \textbf{Samp.} & \textbf{Ratio} \\
\hline
FAM:pcacceleratepro & pup & peexe & 1,749 & 9.5\% \\ FAM:sagent & down. & macro & 2,141 & 9.3\% \\ FAM:dstudio & down. & peexe & 1,255 & 6.2\% \\ FAM:pasnaino & down. & peexe & 613 & 5.9\% \\ FAM:opensupdater & pup & peexe & 2,051 & 4.8\% \\ FAM:mobtes & down. & apk & 967 & 4.6\% \\ FAM:hesv & pup & peexe & 849 & 4.4\% \\ FAM:asacub & infosteal & apk & 833 & 4.1\% \\ UNK:agentino & down. & peexe & 649 & 4.0\% \\ FAM:fakecop & pup & apk & 672 & 3.6\% \\ \hline
\end{tabular}
\caption{Top 10 families (with over 10K samples) sorted by ratio of originally 
FUD samples.
}
\label{tab:fudFamilies}
\end{table}

\ignore{
\begin{table}[t]
\small
\centering
\begin{tabular}{lllr}
\hline
\textbf{Family} & \textbf{Filetype} & \textbf{Class} & \textbf{Samples} \\
\hline
FAM:berbew & peexe & backdoor & 36,919 \\
FAM:virlock & peexe & virus & 16,333 \\
FAM:dinwod & peexe & downloader & 16,122 \\
FAM:pajetbin & peexe & worm & 12,798 \\
FAM:faceliker & javascript & clicker & 12,294 \\
FAM:dridex & peexe & infostealer & 11,006 \\
FAM:smsreg & apk, javascript & pup & 9,210 \\
FAM:cryxos & javascript & rogueware & 9,119 \\
FAM:banload & peexe & miner & 6,731 \\
FAM:stihat & peexe & worm & 6,068 \\
FAM:mansabo & peexe & miner & 5,828 \\
FAM:lamer & peexe & virus & 5,678 \\
UNK:lnkrun & - & - & 5,426 \\ UNK:bulz & peexe & - & 5,394 \\ FAM:salgorea & peexe & downloader & 5,345 \\
\hline
\end{tabular}
\caption{Top 15 families for FUD samples.
}
\label{tab:fudFamilies}
\end{table}
}

Using their last report, 
\avclass outputs a family for 62.5\% of the 600K originally FUD samples, 
which is in line with the overall labeling rate, 
indicating a similar fraction of well-known families among 
originally FUD samples.
However, some families have larger fractions of originally FUD samples, 
and thus are harder to detect.
Table~\ref{tab:fudFamilies} shows the top 10 families 
with at least 10K samples sorted by the ratio of 
originally FUD samples over all family samples.
These include 6 families for Windows, 3 for Android, and 
one family of Microsoft Office macros.
All of them have FUD ratios 6--16 times higher than the 0.59\% average 
over all families with at least 10K samples.

\take{At least 0.3\% of samples submitted to VT are originally FUD, 
i.e., they have zero detections on the first VT observation, 
but later are considered malicious by at least 4 engines.
PDF documents are more likely to be FUD than other filetypes.}

\paragraph{Code signing.}
VT extracts code signatures from multiple filetypes. 
The collected reports contain 13.3M samples (5.6\% of all samples) 
for which VT extracted code signing certificates.
Of the signed samples, 
55.9\% are Android APKs, 
43.4\% are Windows PE files, and
0.7\% are other filetypes such as 
Microsoft Installers (.msi) and patches (.msp), 
Mach-O executables, 
iOS applications, 
Apple image files (.dmg), and 
some archive formats (e.g., .zip, .cab).
PDF is one popular filetype for which VT does not currently 
extract signatures.
91.3\% of all \emph{apk} samples, 
3.7\% \emph{peexe} ,
31.4\% \emph{msi}, and
7.6\% \emph{macho} 
are signed.
APKs have to be signed in order for the Android OS to install them in a 
device. 
The 8.5\% of unsigned APKs is due to 
apps under development being uploaded to VT, 
possibly to check if any AV engine detects them or as part of 
continuous delivery pipelines. 

\platon{Note for me: Report number of VT reports in which we see a
certificate but it is not properly parsed (e.g., misses Subject/Issuer info)}

\take{VT supports the extraction of code signatures for a variety of filetypes,
but only a small fraction (5.6\%) of all feed samples, 
and 3.7\% of the \emph{peexe} samples, have a code signing signature. 
}

\paragraph{Telemetry intersection.}
We intersect the hashes of feed samples with the telemetry. 
However, to make the telemetry query scale 
we had to exclude known benign samples 
using a curated whitelist from the \vendor.
The intersection contains 3.8M samples with at least one detection 
(1.8\% of feed samples with one detection) and 
2.2M (1.2\%) with at least four detections.
The small intersection indicates that the telemetry and the VT file feed 
observe largely disjoint sets of malicious samples.
Prior work has observed that public and commercial threat intelligence feeds 
have small overlap~\cite{tealeaves,bouwman2020different}.
However, those works did not evaluate VirusTotal
or the telemetry of a large \vendor, 
which are arguably the largest datasets for indicators such as file hashes.
Our results show that even the largest datasets are largely disjoint with 
minimal overlap.
This is likely caused by a huge space of malicious samples of which 
each vendor only sees a small portion.

\begin{table}[t]
\small
\centering
\begin{tabular}{llrr}
\hline
\textbf{Family} & \textbf{Class} & \textbf{Devices}  & \textbf{Samples} \\
\hline
FAM:winactivator       & pup  &   2.0M  &  10,871 \\
FAM:tool:utorrent      & pup  &   1.6M  &   1,366 \\
FAM:installcore        & pup  &   1.5M  &  46,758 \\
FAM:webcompanion       & pup  &   1.4M  &   2,569 \\
FAM:dotsetupio         & pup  &   1.1M  &    198  \\
FAM:iobit              & pup  &   898K  &   4,321 \\
FAM:opensupdater       & pup  &   692K  &  14,918 \\
FAM:opencandy          & pup  &   579K  &   9,346 \\
UNK:offercore          & pup  &   555K  &     363 \\
FAM:driverreviver      & pup  &   545K  &     615 \\
\hline
\end{tabular}
\caption{Top 10 families for feed samples in the telemetry.
}
\label{tab:telemetryFamilies}
\end{table}

\ignore{
\begin{table}[t]
\small
\centering
\begin{tabular}{llr}
\hline
\textbf{Family} & \textbf{Class} & \textbf{Samples} \\
\hline
FAM:installcore & pup & 46,189 \\ FAM:neshta & virus & 44,389 \\
FAM:agenttesla & infosteal & 41,246 \\
FAM:emotet & infosteal & 24,680 \\
FAM:razy & infosteal & 22,435 \\ UNK:bulz & - & 20,809 \\
FAM:softcnapp & pup & 20,021 \\
FAM:ekstak & pup & 19,017 \\
FAM:onlinegames & infosteal & 18,879 \\
FAM:berbew & backdoor & 17,136 \\
\hline
\end{tabular}
\caption{Top 10 families for feed samples in the telemetry.
}
\label{tab:telemetryFamilies}
\end{table}
}

Table~\ref{tab:telemetryFamilies} shows the top 10 families for feed samples 
in the telemetry.
We observe the stark difference with the top families in the feed 
(\emph{peexe} families in Table~\ref{tab:families_per_filetype}) 
with no families in common.
All top ten families in the telemetry are PUP while 
the feed's top families are dominated by virus and worm families. 
From the top 10 feed families, \emph{vobfus} and \emph{virlock} 
are the two families ranked highest in the telemetry 
found on 25.9K and 3.3K devices respectively.
The remaining 8 families are ranked below the 1,000th position affecting each
less than 2K devices.
The fact that the distribution of malicious families in the 
VT file feed widely differs from that observed on real devices
is important for works that may try to analyze the malware ecosystem 
using VT. 
While VT is a great source of malicious samples, its file distribution is 
biased towards families contributed more often 
(e.g., those with high polymorphism), 
and thus it does not capture impact on real devices.

During our collection period, there are 11.9M samples detected as 
malicious by the \vendor, 
but not observed in the VT file feed.
These files are either 
never submitted to VT or
their last VT report was before our collection start.
Quantifying this requires querying the 11.9M files to VT
which due to API restrictions is not possible.
Instead, we estimate these figures by querying a subset of 1M randomly 
selected hashes.
Only 10.9\% of those are known to VT, while 89.1\% have never been submitted.
This shows that security vendors may only share a fraction of their malicious 
samples with VT. 
Sharing decisions are taken by other \vendor departments and 
are transparent to us.

From the telemetry, we also obtain the telemetry first seen timestamp, 
which is the earliest time, within our collection period, 
a feed sample was queried by an endpoint to obtain its reputation. 
This is an upper bound on 
the earliest time the \vendor observed the sample.
For example, a sample first seen by the \vendor in 2010 may  
appear in the telemetry subset we analyze as first queried on 
December 22nd, 2020.
We calculate the delay to observe a sample 
as the VT first seen timestamp minus 
the telemetry first seen timestamp, 
but only for the 2.1M new samples (from VT perspective) 
with at least one detection and in the intersection with the telemetry.
Of those 2.1M samples, 
2.5M (61\%) are first observed by the telemetry
(i.e., positive difference)
while 816K (39\%) are first observed by VT
(i.e., negative difference).
The median delay for VT to observe the sample is 4.4 hours.
Thus, real devices observe the sample a few hours earlier than VT. 
However, the mean delay is 21 days because 12\% of new samples  
are first submitted to VT at least 3 months after they appear in the telemetry, 
compared to 3\% being observed by VT 3 months earlier than in the telemetry. 
It is important to note that since the telemetry first seen is an upper 
bound for the \vendor first seen, 
the VT delay may be actually larger.

\take{The telemetry and VT file feed observe largely disjoint sets of 
malicious samples 
(1.2\%--1.8\% of feed samples in common).
Thus, even the largest security vendors only see a 
small portion of the whole space of malicious samples.
The family distribution in the VT file feed significantly 
differs from that observed in the telemetry indicating that VT 
may not be a good source to infer family impact on real users.
New samples in both datasets are seen a median of 4.4 hours earlier 
by the devices. 
Thus, new malicious samples in the feed are quite recent.
Furthermore, 39\% of new samples are observed by VT before they are seen 
in user devices, so VT may provide useful early alerts to {\vendor}s.
}
\section{Clustering}
\label{sec:clustering}

The goal of the clustering is to group samples that belong to the same family.
We require highly scalable clustering approaches, 
as we need to cluster 235.7M samples, 
more than five times the 42M samples clustered in the 
largest-to-date work~\cite{avclass2}.
Another property we desire is that the clustering produces 
high precision clusters, despite a potentially low recall. 
High precision means the clusters have high purity, 
i.e., the vast majority of samples in the cluster belong to the same family. 
Low recall means that a family may be split into a potentially large number 
of clusters. 
The reason to favor high precision is that it enables the analyst 
to gain an understanding of the cluster 
by analyzing one (or a small number) of samples in the cluster, 
thus reducing the number of threats that need analysis.
Similarly, given a sample of interest, its cluster can be used 
to infer properties of the sample. 
For example, 
inferring that an undetected sample is malicious because
it belongs to a cluster where most samples are detected.
Such generalization is problematic when the precision is low.

\begin{table}[t]
\small
\centering
\begin{tabular}{l|r|rrrr}
\cline{3-6}
\multicolumn{2}{c}{} & \multicolumn{4}{c}{\textbf{Values}} \\
\hline
\textbf{Feature}  & \textbf{Samples} & \textbf{All} & \textbf{peexe} & \textbf{apk} & \textbf{other}\\
\hline
tlsh 	          & 235.6M &  204.3M  &  134.0M & 7.6M  &  62.7M \\
vhash           & 220.1M &   17.4M  &    6.4M &  1.3M &  9.7M  \\
authentihash    & 155.3M &  152.9M  &  152.9M &    0  &    0  \\ avc2\_family    & 151.7M &   74.4K  &   62.4K &  2.8K &   22.4K  \\
imphash         & 149.7M &    8.3M  &    8.3M &    0  &    0  \\ richpe\_hash    &  65.2M &    2.1M  &    2.1M &    0  &    0  \\ icon\_hash      &  60.5M &   15.6M  &    1.9M & 2.2M  &  11.5M \\
cert\_thumb.    &  13.3M &    1.7M	&  168.7K & 1.5M  & 10.6K		\\
pkg\_name   &   7.4M &    2.8M  &      0  &  2.8M &     0  \\
\hline
\end{tabular}
\caption{For the 10 clustering features, number of samples with non-NULL 
feature value, distinct feature values, and split of feature values 
by filetype.
}
\label{tab:featurecount}
\end{table}

\paragraph{Clustering features.}
To compute sample similarity, we are limited to the features 
we extract from the VT reports.
From the \nfeatures features in Table~\ref{tab:features} we examine 9 features 
that can identify similar samples.
Table~\ref{tab:featurecount} shows for each feature 
the number of samples that have a non-NULL feature value, 
the number of distinct feature values, and the split of feature values 
for \emph{peexe}, \emph{apk}, and other filetypes.
Two features are available in most samples
regardless of filetype:
\feature{tlsh} is extracted for every sample save 
a handful of errors (99.9\%) and 
\feature{vhash} for most filetypes with some exceptions such as 
images (93.4\%).
Other features available for multiple filetypes are 
the \avclass family (63.5\%), 
the icon's MD5 (25.7\%), and
the certificate thumbprint (5.6\%).
Three features are specific to \emph{peexe} samples: 
\emph{authentihash} and \emph{imphash} are 
both available for 96\% of \emph{peexe} samples, 
while 42\% of \emph{peexe} samples have the optional Rich PE 
header.
The package name is specific to \emph{apk} samples, 
being available in 90\% of those.

\paragraph{Clustering approaches.}
We evaluate three clustering approaches that do not need to define the 
number of clusters a priori. 
The first approach is what we call \emph{feature value grouping} (FVG), 
a simple grouping of samples using equality on the 
values of a single feature, 
e.g., all samples with the same \feature{vhash} value form a cluster. 
Samples that lack the used feature 
(e.g., those with a NULL \feature{vhash})
are placed in a singleton cluster by themselves.
To cluster using FVG the feature vectors are sorted by the triple
(feature\_value, sample\_hash, scan\_date).
A change in feature value 
indicates a new cluster.
FVG is an extremely scalable approach. 
We evaluate FVG with different features to understand which ones
provide high precision, and what recall those features achieve.

The second approach is hierarchical agglomerative clustering (HAC) 
with single linkage. 
Our distance function is a weighted average of the feature distances 
where the feature distance is 
boolean for cryptographic hashes and \feature{vhash}, and
Hamming distance for \feature{tlsh}.
To avoid biasing the clustering towards specific features, 
we apply equal weight to all features.
When a sample misses one feature (e.g., has no \feature{vhash}), 
we exclude the feature from the weighted average.
We also evaluate different values for the distance threshold parameter,
which defines the cluster diameter.
The main limitation of hierarchical clustering is that it requires 
computing a quadratic number of comparisons. 

The third approach is the recent 
\emph{threshold-based hierarchical agglomerative clustering} 
(HAC-T)~\cite{oliver2020hac},
an scalable approach to cluster 
samples using their TLSH hash value. 
It uses vantage trees to achieve $\mathcal{O}(n\log{}n)$ comparisons 
in contrast to the typical $\mathcal{O}(n^2)$ in hierarchical clustering.
It has been shown to produce high purity clusters, 
has been applied to cluster 10M samples, and 
a Python version has been open sourced~\cite{tlshRepo}.
We also evaluate a variant called \emph{HAC-T-opt} 
that provides optimal clusters at some performance cost~\cite{oliverfast}.
Both HAC-T and HAC-T-opt use a cluster distance ($CDist$) parameter, 
which we set to the recommended value of 30~\cite{oliverfast}.

\begin{table}[t]
\small
\centering
\begin{tabular}{|l|l|r|r|c|l|}
\hline
  \textbf{Dataset}
  &\textbf{Plat.}
  &\textbf{Samples}
  &\textbf{Fam.}
  &\textbf{Collection} \\
\hline
AMD~\cite{amd}                        & And & 24,551    & 71  & 11/2010 - 03/2016\\ Malicia~\cite{malicia}                & Win & 9,908     & 52  & 03/2012 - 02/2013\\
\hline
\end{tabular}
\caption{Ground truth datasets used to evaluate clustering.}
\label{tab:gt}
\end{table}

\begin{table}[t]
\small
\centering
\begin{tabular}{llrrrr}
\hline
\textbf{Feature} & \textbf{Algor.} & \textbf{Clust.} & \textbf{Prec.} & \textbf{Recall} & \textbf{F1} \\ \hline
authentihash  & fvg & 9,909 & 100\%  &  0.5\% &  1.1\% \\ avc2\_family  & fvg &   284 & 97.0\% & 75.4\% & 84.8\% \\ cert\_thumb.  & fvg & 9,410 & 100\%  &  1.8\% &  3.5\% \\ icon\_hash    & fvg & 9,766 & 99.9\% &  1.0\% &  1.9\% \\ imphash       & fvg & 1,843 & 99.7\% &  5.7\% & 10.7\% \\ richpe\_hash  & fvg & 9,899 & 100\%  &  0.6\% &  1.2\% \\ vhash         & fvg &   900 & 98.8\% & 12.8\% & 22.7\% \\ \hline
tlsh & hact-opt	& 3,772 & 99.9\% & 6.2\% & 11.8\% \\ tlsh & hact & 3,899 & 99.9\% & 3.8\% & 7.3\% \\	\hline
\end{tabular}
\caption{Clustering accuracy on Malicia dataset.
Samples is the percentage of samples with a non-NULL feature value.
}
\label{tab:malicia_clustering}
\end{table}

\begin{table}[t]
\small
\centering
\begin{tabular}{llrrrr}
\hline
\textbf{Feature} & \textbf{Algor.} & \textbf{Clust.} & \textbf{Prec.} & \textbf{Recall} & \textbf{F1} \\ \hline
avc2\_family       & fvg &  407  & 95.4\% & 89.0\% & 92.1\% \\ cert\_thumb.   & fvg & 8,566  &   90.1\% & 15.9\% & 27.0\% \\ icon\_hash     & fvg & 17,812  &  98.7\% &  9.7\% & 17.6\% \\ pkg\_name      & fvg & 17,430  &  99.6\% &  9.2\% & 16.8\% \\ vhash          & fvg &  7,478  &  94.5\% & 20.6\% & 33.8\% \\ \hline
tlsh & hact-opt	& 17,515 & 98.3\% 	& 7.6\% 	& 14.1\%	\\ tlsh & hact & 17,884 & 98.4\%	& 6.9\%	& 12.9\%	\\ \hline
\end{tabular}
\caption{Clustering accuracy on AMD dataset.
}
\label{tab:amd_clustering}
\end{table}

\paragraph{Accuracy evaluation.}
We measure clustering accuracy 
using two ground truth datasets summarized in Table~\ref{tab:gt}. 
The Malicia dataset contains nearly 10K Windows 
malware collected from drive-by downloads~\cite{malicia}. 
The AMD dataset contains 24K Android samples~\cite{amd}.
Those are the two largest family-labeled public datasets we could find.
For Malicia, we requested VT a re-scan of all samples because 
most samples did not have a TLSH hash value, 
as they had been scanned before VT introduced TLSH. 
For AMD, a re-scan was not needed, 
but we still collected the latest reports.

Tables~\ref{tab:malicia_clustering} and~\ref{tab:amd_clustering} 
summarize the clustering accuracy on the two datasets.
For FVG, on AMD we exclude PE-specific features 
(\feature{authentihash}, \feature{imphash}, \feature{richpe\_hash}) and
on Malicia APK-specific features (\feature{pkg\_name}).
For HAC we test different combinations of features and distance thresholds 
achieving a best result of 24.0\% F1 score using a distance threshold of 0.8 
when not including the \feature{avc2\_family} feature and up to 
83.8\% F1 score when including that feature.
\juan{Add HAC results to tables}
Samples with a NULL feature value are placed in a singleton cluster, 
thus increasing the number of clusters and lowering the recall. 
Thus, features only available for a small fraction of samples such as
\feature{richpe\_hash} (1.0\%) and icon hash (6.1\%) tend to have 
very low recall and F1 score.
Other features like \feature{authentihash} are widely available (98.1\%), 
but rarely identify similar samples. 
In particular, \feature{authentihash} only groups samples that are the 
same malware variant, but signed with different certificate chains.
Overall, all features achieve very high precision ranging 97.0\%--100\% 
except \feature{cert\_thumbprint} on AMD with 90.1\%.
However, recall widely differs among features.
Results are consistently better on AMD and the best feature is 
\feature{avc2\_family},
which achieves 84.8\% F1 score on Malicia and 92.1\% on AMD, 
followed by \feature{vhash} (22.7\%--33.8\%), 
\feature{cert\_thumbprint} (27.0\%), and 
TLSH using HAC-T-opt (11.8\%--14.1\%).
It is important to note that the \avclass family is the only 
feature that may improve over time since AV labels may be refined 
as AVs improve their detection signatures.
Since the Malicia dataset is from 2013 and AMD from 2016
the results may overestimate the accuracy of \avclass on newer samples.
Still, Section~\ref{sec:vtfeed} showed that 62.3\% of samples 
can be labeled on first sight.

\paragraph{Scalability.}
While all three clustering approaches successfully run on the GT datasets,
those datasets are pretty small. 
To evaluate scalability, we run the clustering approaches on different 
days of the feed with increasingly larger number of samples. 
We set an upper limit of 24 hours for the clustering to complete, 
so that the clustering can handle the daily influx of feed samples.
We run the clustering on a server with an 
Intel(R) Xeon(R) CPU E5-2660 v3 @ 2.60GHz with 40 cores and 128GB RAM.
The lowest number of daily samples corresponds to the first day with 620K samples.
On this day, the hierarchical clustering does not complete in the allocated 
24h showing that the quadratic number of comparisons it requires does not 
scale to the daily number of feed samples. 
On this first day, HAC-T takes 9.7 hours and FVG-vhash
takes less than one hour
(49.6 minutes for sorting and 2.6 minutes for grouping).
Next, we run the clustering on an average day with 1.4M samples where HAC-T takes 22.4 hours and FVG-vhash again less than one hour.
When run on the day with the highest number of samples (2.2M) 
HAC-T does not terminate in the allocated 24 hours, however FVG-vhash 
still takes less than one hour. 
The original HAC-T paper mentions that they are able to cluster 10M samples 
in 2.2h using commodity hardware similar to what we use. 
One possible explanation for the discrepancy is that the we use 
the HAC-T open source Python implementation, 
while the original paper used a more efficient C version.
\juan{Think and refine}

\begin{table*}[t]
\small
\centering
\begin{tabular}{|l|c|rr|rr|rrrr|rr|}
\cline{3-12}
\multicolumn{2}{c|}{} & \multicolumn{2}{c|}{\textbf{Clusters}} & \multicolumn{2}{c|}{\textbf{Singletons}} & \multicolumn{4}{c|}{\textbf{Non-NULL clusters}} & \multicolumn{2}{c|}{\textbf{Runtime (h)}} \\
\hline
\textbf{Feature} & \textbf{Filetype} & \textbf{All} & \textbf{Non-NULL} & \textbf{All} & \textbf{Non-NULL} & \textbf{Max.} & \textbf{Mean} & \textbf{Med.} & \textbf{Std} & \textbf{Sort} & \textbf{Group}\\
\hline
cert\_thumb.	& All	& 224.1M	&  1.7M	&  223.7M	&  1.2M	&  1.1M &  7.8	& 1.0	&     959.4 & 5.0 & 3.5 \\
avc2\_family	& All	&  84.1M 	& 74.4K &  84.1M   	& 41.5K & 19.4M &  2.0 	& 1.0 	& 100,287.4 & 3.9 & 10.3 \\
vhash			& All	&  33.2M	& 17.4M	&   29.2M	& 13.4M	&  3.3M & 12.7	& 1.0	&   2,138.0 & 4.2 & \kevin{Rerun}11.1 \\
imphash			& Peexe	&  14.1M   	&  8.3M	&   11.6M  	&  5.7M &  6.0M & 18.1	& 1.0 	&   5,537.1 & 5.9 & 2.3 \\

\hline
\end{tabular}
\caption{Clustering results for FVG with different features on the whole dataset of 235.7M samples (except for imphash).}
\label{tab:clustering}
\end{table*}

\paragraph{Clustering the whole dataset.}
Since FVG is the only clustering approach that scales to the largest day, 
we apply it to cluster the 235.7M feed samples.
We use four features: 
the three features that support multiple filetypes and 
achieved best F1 score on both Malicia and AMD 
(\emph{avc2\_family}, \emph{vhash}, \emph{cert\_thumprint}) plus 
\emph{imphash} that is specific to Windows samples, but performs 
well on those (i.e., fourth F1 on Malicia).
The clustering took between 8.2 hours for \emph{imphash} up to 
15.3 hours for \emph{vhash}.
Table~\ref{tab:clustering} summarizes the results.
It shows the total number of clusters produced, as well as 
the number of clusters excluding singleton clusters created 
for samples that have a NULL feature value.
Similarly, it lists the total number of singleton clusters and those 
for samples with a non-NULL feature value, 
which matches the number of feature values in Table~\ref{tab:featurecount}.
For example, FVG-vhash produces 33.2M clusters, 
of which 29.2M (88\%) are singletons and 4.0M (12\%) contain multiple samples. 
Of those singletons, 15.8M (54\%) correspond to samples without a vhash value, 
while 13.4M (46\%) correspond to samples with a vhash value not shared with 
any other sample. 
The two features that are able to cluster more samples are \emph{imphash} 
with a mean non-NULL cluster size of 18.1 followed by \emph{vhash} (12.7). 
Still, all features have a median of one, showing that their distribution 
has at least 50\% of singletons for samples with a non-NULL feature value.
Depending on the application, analysts may want to first filter by filetype 
before clustering. 
We opted to include all filetypes in the clustering
(for those features supporting multiple filetypes) 
to push the clustering scalability to the limit.

\section{Detecting Not-so-Benign Samples}
\label{sec:hunting}

In this section we leverage the produced clusters for  
detecting samples thought to be benign (i.e., with zero detections) 
that may really be malicious because 
they belong to a cluster where most samples are detected as malicious. 

\begin{table}[t]
\small
\centering
\begin{tabular}{|l|rr|rr|rr|}
\cline{2-7}
\multicolumn{1}{c|}{} & \multicolumn{2}{c|}{\textbf{Vhash}} & \multicolumn{2}{c|}{\textbf{Imphash}} & \multicolumn{2}{c|}{\textbf{Thumbprint}} \\
\hline
\textbf{Cluster Type} & \textbf{r1} & \textbf{r4} & \textbf{r1} & \textbf{r4} & \textbf{r1} & \textbf{r4} \\
\hline
Fully malic.  & 73\% & 46\% & 89\% & 81\% & 50\% & 19\% \\
Fully benign  & 19\% & 44\% &  6\% & 14\% & 23\% & 65\% \\
Mixed         &  8\% & 10\% &  5\% &  5\% & 27\% & 15\% \\
\hline
Mal. majority &  5\% &  6\% &  3\% &  3\% & 17\% &  8\% \\
\hline
\end{tabular}
\caption{Fraction non-singleton FVG clusters classified as 
fully malicious, fully benign, mixed, or malicious majority.}
\label{tab:undetected}
\end{table}

\ignore{
\begin{table}[t]
\small
\centering
\begin{tabular}{|l|rr|rr|rr|}
\cline{2-7}
\multicolumn{1}{c|}{} & \multicolumn{2}{c|}{\textbf{Vhash}} & \multicolumn{2}{c|}{\textbf{Imphash}} & \multicolumn{2}{c|}{\textbf{Thumbprint}} \\
\hline
\textbf{Cluster Type} & \textbf{vt1} & \textbf{vt4} & \textbf{vt1} & \textbf{vt4} & \textbf{vt1} & \textbf{vt4} \\
\hline
Fully malic.	& 72.6\% & 46.5\% & 89.5\% & 81.1\% & 49.8\% & 19.3\% \\
Fully benign	& 19.3\% & 43.9\% & 5.7\% & 14.2\% & 23.1\% & 65.3\% \\
Mixed	& 8.1\% & 9.7\% & 4.8\% & 4.8\% & 27.0\% & 15.4\% \\
\hline
Mal. majority	& 5.5\% & 6.5\% & 3.5\% & 3.1\% & 16.7\% & 8.3\% \\
\hline
\end{tabular}
\caption{Classification of non-singleton FVG clusters with different features on whole feed.}
\label{tab:undetected}
\end{table}
}

For each cluster that is not a singleton, we compute two ratios: 
the number of samples with at least one detection over all cluster samples 
(r1) and 
the number of samples with at least four detections over all cluster samples
(r4).
Table~\ref{tab:undetected} shows the fraction of non-singleton FVD 
clusters that belong to four classes:
\emph{fully malicious} clusters that have an rx of 1 (where x is 1 or 4);
\emph{fully benign} clusters with an rx of 0; and
mixed clusters that contain a mixture of detected and undetected samples
(i.e., 0 < rx < 1).
For example, of the 4.0M non-singleton FVD-vhash clusters, 46\% contain only samples with at least four detections 
(i.e., the cluster is fully malicious) and 
44\% have no samples with at least four detections
(i.e., the cluster is benign).
Of the remaining 10\% non-singleton clusters 
29,786 have at least half of the samples detected by at least four AV engines.
Those 29K clusters contain 190K samples with no detections, which could be identified as malicious 
using their context.
We illustrate such clusters with two examples.

\juan{Use FUD samples as some kind of ground truth?}

\paragraph{vhash:7a41a7dd4a319c77194ec2a5f6c78aa1.}
This cluster has 13,172 APK samples. 
Of those, 13,167 (99.96\%) are detected by at least one AV engine and
13,152 (99.85\%) are detected by more than 4 AV engines.
Thus, the cluster is clearly malicious, but there are 5 samples 
that no AV engine detects as malicious.
Those 5 samples can be reported to an 
analyst to check if they are indeed malicious and why they may be undetected.
Similarly, there are an additional 15 samples with a low number of detections
(i.e., less than four) that may also be of interest to an analyst.

\paragraph{thumb:0069be5da75d35577a5b1b02810571fc72d43162.}
This cluster has 12 \emph{peexe} samples, 
all with an invalid certificate chain that contain the same self-signed 
leaf certificate with subject ``ZIRPER\textbackslash zirpe''.
There are 11 (91.67\%) samples detected by at least 4 AV engines, 
but one sample has zero detections. 
Samples in the cluster have 4 different vhash values, 
no imphash values extracted by VT, and three filenames: 
AntiCheatS.exe (7 samples), 
AntiCheat.exe (4), and 
AntiCheatUpdater.exe (1). 
The undetected sample has a vhash not shared with other samples, 
but it has the most popular filename.

\section{Related Work}
\label{sec:related}

Most related is the work by Ugarte-Pedrero et al.~\cite{ugarte2019close} 
that analyzes one day of the samples a large \vendor
collects through multiple sources. 
They analyze 172K PE executables and cluster 58K using HAC.
In contrast, we examine one year of a file feed with 235M samples of 
multiple filetypes, 
compare it to the telemetry of a large \vendor, and 
analyze scalable approaches to cluster the 235M samples.
Other works have performed large scale longitudinal analysis on
Windows~\cite{bayer2009view,lever2017lustrum}, 
Android~\cite{lindorfer2014andrubis,suarez2020eight}, and 
Linux~\cite{cozzi2018understanding,alrawi2021circle} malware.
In contrast, our work examines malware for multiple platforms 
including Windows, Android, Linux, macOS, Microsoft Office macros, 
PDF documents, and Web content.

Extensive research has addressed the problem 
of clustering malware samples into families~\cite{bayer2009scalable,perdisci2010behavioral,bitshred,mutantx,firma,li2015experimental,avclass,faridi2018performance,avclass2,oliver2020hac}. 
However, none of those works has attempted to cluster as large a 
dataset as ours, more than five times the 42M samples clustered in the 
largest-to-date work~\cite{avclass2}.
Other related works address the problem of malware triage, 
i.e., how to prioritize samples for analysis~\cite{mast,sigmal,vilo,buyukkayhaa2017n,laurenza2017malware,spotlight}.
Most related among these is 
Spotlight~\cite{spotlight}, which takes as input malware,
applies supervised classifiers to identify samples from well-known families, 
clusters the remaining unlabeled samples,
and ranks the clusters with an application-dependant scoring function.
In contrast, we use \avclass to identify samples belonging to well-known 
families and do not rank clusters, but identify 
clusters of interest that contain a majority of malware together with 
undetected samples.

\section{Conclusions}
\label{sec:conclusions}

We have characterized the VirusTotal feed 
by analyzing 328M reports for 235M samples
collected from the feed during one year, and have compared the feed 
with the telemetry of a large \vendor.
Among others, we show that despite having a volume 17 times
lower than the telemetry,
the feed observes 8 times more malware.
The feed is fresh; 89\% samples are new and
samples appear a median of 4.4 hours after they are
seen in user devices.
The feed is also diverse containing 
4.9K families with at least 100 samples.
However, the diversity is largely due to Windows and Android families.
On the negative side, the family distribution
significantly differs from that observed in the telemetry indicating that
VT may not be a good source to infer family impact on real users.
We have also evaluated three clustering approaches with the goal of 
producing high precision clusters that scale to the 235M samples.
Our results show that a simple grouping by feature value 
can produce clusters with 0.90--0.99 precision for selected features, 
and can cluster the 235M samples in 15 hours.
Using FVG-vhash, we identify 29K clusters with a majority of
samples detected by multiple AV engines,
containing 190K likely malicious samples with zero detections.

\section*{Acknowledgments}
This work has been partially supported by the SCUM Project 
(RTI2018-102043-B-I00) MCIN/AEI/10.13039/501100011033/ERDF 
A way of making Europe. 
This work has been partially supported by the Madrid regional government 
as part of the program S2018/TCS-4339 (BLOQUES-CM), 
co-funded by EIE Funds of the European Union.
Any opinions, findings, and conclusions or recommendations in 
this material are those of the authors or originators, and 
do not necessarily reflect the views of the sponsors.

\appendix
\section*{Appendix}

\newpage
\begin{table}[t]
	\footnotesize
	\centering
	\begin{tabular}{lllr}
		\hline
		\textbf{Filetype} & \textbf{Family} & \textbf{Class} & \textbf{Samples}\\
		\hline
		
javascript
& FAM:faceliker   & clicker &   2,288,894 \\ & FAM:facelike    & pup &   952,180 \\ & FAM:coinhive    & miner &   766,087 \\ & FAM:cryxos  & - &   744,894 \\ & FAM:smsreg  & pup &   415,669 \\ & UNK:gnaeus  & - &   400,570 \\ & FAM:fakejquery  & downloader &   330,792 \\ & UNK:hidelink    & - &   210,306 \\ & UNK:expkit  & - &   96,433 \\ & UNK:agentwdcr   & - &   87,101 \\ \hline
html
& UNK:refresh & - &   882,026 \\ & FAM:cryxos  & - &   363,821 \\ & FAM:faceliker   & clicker &   312,563 \\ & FAM:smsreg  & pup &   201,253 \\ & UNK:redir   & downloader &   200,926 \\ & FAM:coinhive    & miner &   152,968 \\ & UNK:generickdz  & pup &   121,975 \\ & UNK:pushnotif   & - &   120,085 \\ & FAM:ramnit  & virus &   80,044 \\ & UNK:fklr    & rogueware &   79,353 \\ \hline
pdf
& UNK:fakeauthent & phishing &   194,963 \\ & UNK:minerva & phishing &   15,527 \\ & FAM:pdfka   & exploit &   13,618 \\ & UNK:pidief  & - &   6,319 \\ & FAM:alien   & downloader &   6,137 \\ & UNK:gorilla & phishing &   4,749 \\ & UNK:talu    & phishing &   2,379 \\ & UNK:gerphish    & phishing &   1,558 \\ & UNK:urlmal  & phishing &   1,469 \\ & FAM:rozena  & pup &   839 \\ 
		\hline
		
	\end{tabular}
	\caption{Other Top 10 families .}
	\label{tab:other_families_per_filetype}
\end{table}
\begin{table}[t]
	\footnotesize
	\centering
	\begin{tabular}{lllr}
		\hline
		\textbf{Filetype} & \textbf{Family} & \textbf{Class} & \textbf{Samples}\\
		\hline
		peexe		
& FAM:berbew	& backdoor &	19,371,273 \\ & FAM:dinwod	& downloader &	9,398,314 \\ & FAM:virlock	& virus &	7,921,534 \\ & FAM:pajetbin	& worm &	7,164,373 \\ & FAM:sivis	& virus &	6,222,693 \\ & FAM:lamer	& virus &	4,074,441 \\ & FAM:salgorea	& downloader &	3,737,865 \\ & FAM:vobfus	& worm &	3,415,996 \\ & FAM:drolnux	& worm &	2,858,975 \\ & FAM:griptolo	& worm &	2,407,104 \\ 
		\hline
		apk		& 
 FAM:smsreg	& pup &	616,406 \\ & FAM:ewind	& pup:adware &	430,531 \\ & FAM:hiddad	& pup:adware &	219,577 \\ & FAM:fakeadblocker	& pup:adware &	82,715 \\ & FAM:adlibrary:airpush	& pup:adware &	80,704 \\ & FAM:adlibrary:revmob	& pup:adware &	78,495 \\ & FAM:dowgin	& pup:adware &	68,522 \\ & FAM:dnotua	& pup &	65,330 \\ & FAM:kuguo	& pup:adware &	63,262 \\ & FAM:mobidash	& pup:adware &	40,016 \\ 
		\hline
		elf		&
FAM:xorddos	& ddos &	287,631 \\ & FAM:mirai	& backoor &	163,525 \\ & FAM:mirai:gafgyt	& backoor &	59,348 \\ & FAM:tsunami	& backoor &	3,381 \\ & FAM:mirai:hajime	& downloader &	2,499 \\ & FAM:mirai:mozi	& backdoor &	1,996 \\ & FAM:setag	& backdoor &	1,454 \\ & FAM:dofloo	& backdoor &	890 \\ & FAM:fakecop	& pup &	805 \\ & FAM:ladvix	& virus &	580 \\ 		\hline
		macho	&
FAM:flashback	& downloader &	33,087 \\ & FAM:mackontrol	& backdoor &	15,459 \\ & FAM:mackeeper	& pup &	15,017 \\ & FAM:evilquest	& ransomware &	7,070 \\ & FAM:cimpli	& pup:adware &	5,444 \\ & FAM:gt32supportgeeks	& pup &	3,453 \\ & FAM:genieo	& pup:adware &	3,339 \\ & FAM:bundlore	& pup:adware &	3,142 \\ & FAM:installcore	& pup:adware &	1,543 \\ & UNK:fplayer	& pup:adware &	905 \\ 		
		\hline
		doc		&
FAM:emotet	& infosteal &	24,643 \\ & FAM:valyria	& downloader &	10,182 \\ & FAM:thus	& virus &	4,917 \\ & FAM:sagent	& downloader &	4,717 \\ & FAM:donoff	& downloader &	2,437 \\ & FAM:sload	& downloader &	1,856 \\ & FAM:marker	& virus &	1,326 \\ & FAM:logan	& downloader &	1,135 \\ & FAM:sdrop	& downloader &	906 \\ & FAM:alien	& downloader &	846 \\ 		
		\hline
		xls		&
UNK:sneaky	& downloader &	23,521 \\ & FAM:qbot	& backdoor &	22,416 \\ & FAM:squirrelwaffle	& downloader &	18,230 \\ & FAM:zloader	& downloader &	12,371 \\ & FAM:sload	& downloader &	9,067 \\ & FAM:sagent	& downloader &	8,581 \\ & FAM:valyria	& downloader &	6,074 \\ & UNK:encdoc	& downloader &	5,703 \\ & FAM:icedid	& downloader &	4,237 \\ & FAM:laroux	& virus &	2,983 \\ 		\hline
	\end{tabular}
	\caption{Top 10 families for executable and macro filetypes.}
	\label{tab:families_per_filetype}
\end{table}

\ignore{
	\subsection{FileType distributions}

\begin{table}[t]
\small
\centering
\begin{tabular}{llrr}
\hline
\textbf{File type} & \textbf{Samples} & \textbf{Samples Perc} \\
\hline
 Win32 EXE                        & 89098893  & 66.3 \\
 TXT                              & 8317302   & 6.2 \\
 ZIP                              & 7454689   & 5.5 \\
 Win32 DLL                        & 7164346   & 5.3 \\
 Win64 EXE                        & 6750548   & 5.0 \\
 PDF                              & 4425774   & 3.3 \\
 GZIP                             & 4274656   & 3.2 \\
 HTML                             & 1631580   & 1.2 \\
 DEX                              & 1466510   & 1.1 \\
 LNK                              & 963715    & 0.7 \\
 Win64 DLL                        & 471961    & 0.4 \\
 ELF executable                   & 362827    & 0.3 \\
 XLS                              & 344318    & 0.3 \\
 FPX                              & 160149    & 0.1 \\
 XLSM                             & 152512    & 0.1 \\
 RAR                              & 132258    & 0.1 \\
 DOC                              & 114146    & 0.1 \\
 PHP                              & 85546     & 0.1 \\
 DOCM                             & 59004     & 0.0 \\
 XLSB                             & 35166     & 0.0 \\
 ELF shared library               & 34763     & 0.0 \\
 Mach-O executable                & 32593     & 0.0 \\
 Mach-O fat binary executable     & 28429     & 0.0 \\
 DOCX                             & 28158     & 0.0 \\
 RTF                              & 25307     & 0.0 \\
 XML                              & 23873     & 0.0 \\
 ISO                              & 22875     & 0.0 \\
 DOS EXE                          & 16286     & 0.0 \\
 PNG                              & 14062     & 0.0 \\
 XLSX                             & 12047     & 0.0 \\
 JSON                             & 8649      & 0.0 \\
 bash script                      & 8249      & 0.0 \\
 Mach-O dynamic link library      & 7663      & 0.0 \\
 BZ2                              & 6480      & 0.0 \\
 sh script                        & 5296      & 0.0 \\
 PPT                              & 4598      & 0.0 \\
 TAR                              & 3872      & 0.0 \\
 MP3                              & 3710      & 0.0 \\
 DOTM                             & 3398      & 0.0 \\
 JPEG                             & 2832      & 0.0 \\
 PPTX                             & 2703      & 0.0 \\
 XLAM                             & 2334      & 0.0 \\
 Mach-O dynamic bound bundle      & 1972      & 0.0 \\
 GIF                              & 1745      & 0.0 \\
 perl script                      & 1708      & 0.0 \\
 SWF                              & 1604      & 0.0 \\
 env script                       & 1098      & 0.0 \\
 Win16 EXE                        & 980       & 0.0 \\
 python script                    & 955       & 0.0 \\
 CHM                              & 930       & 0.0 \\
\hline
\end{tabular}
\caption{Top 50 ExifTool file types.}
\label{tab:exif_ftypes}
\end{table}

\begin{table}[t]
\small
\centering
\begin{tabular}{llrr}
\hline
\textbf{File type} & \textbf{Samples} & \textbf{Perc} \\
\hline
 Win32 Executable MS Visual C++ (generic)                           & 20396086  & 15.2 \\
 Win32 Dynamic Link Library (generic)                               & 11992911  & 8.9 \\
 DOS Executable Generic                                             & 8603293   & 6.4 \\
 Win16 NE executable (generic)                                      & 8524541   & 6.3 \\
 Win64 Executable (generic)                                         & 8210334   & 6.1 \\
 Windows Control Panel Item (generic)                               & 7816530   & 5.8 \\
 Win32 Executable (generic)                                         & 5682061   & 4.2 \\
 UPX compressed Win64 Executable                                    & 4971132   & 3.7 \\
 Android Package                                                    & 4582210   & 3.4 \\
 Adobe Portable Document Format                                     & 4427435   & 3.3 \\
 GZipped data                                                       & 4275080   & 3.2 \\
 Microsoft Visual C++ compiled executable (generic)                 & 4171984   & 3.1 \\
 UPX compressed Win32 Executable                                    & 3671162   & 2.7 \\
 Generic CIL Executable (.NET  Mono  etc.)                          & 3576348   & 2.7 \\
 Win32 Executable Microsoft Visual Basic 6                          & 2168482   & 1.6 \\
 OS/2 Executable (generic)                                          & 1804692   & 1.3 \\
 ZIP compressed archive                                             & 1675310   & 1.2 \\
 Dalvik Dex class                                                   & 1473616   & 1.1 \\
 Inno Setup installer                                               & 1461710   & 1.1 \\
 InstallShield setup                                                & 1429299   & 1.1 \\
 Windows screen saver                                               & 1113420   & 0.8 \\
 Win32 Executable Delphi generic                                    & 1110326   & 0.8 \\
 NSIS - Nullsoft Scriptable Install System                          & 1109528   & 0.8 \\
 Windows Shortcut                                                   & 964124    & 0.7 \\
 Win32 EXE PECompact compressed (generic)                           & 961121    & 0.7 \\
 HyperText Markup Language                                          & 829776    & 0.6 \\
 Java Archive                                                       & 827683    & 0.6 \\
 Win32 Executable Borland Delphi 7                                  & 703440    & 0.5 \\
 Digital Micrograph Script                                          & 661216    & 0.5 \\
 HyperText Markup Language with DOCTYPE                             & 598398    & 0.4 \\
 file seems to be plain text/ASCII                                  & 583665    & 0.4 \\
 DOS Borland compiled Executable (generic)                          & 552621    & 0.4 \\
 ASPack compressed Win32 Executable (generic)                       & 488147    & 0.4 \\
 Win32 Executable kkrunchy compressed                               & 473200    & 0.4 \\
 WinRAR Self Extracting archive (4.x-5.x)                           & 316260    & 0.2 \\
 ELF Executable and Linkable format (Linux)                         & 264760    & 0.2 \\
 Extensible Firmware Interface x86-64 application                   & 239904    & 0.2 \\
 Microsoft Excel sheet                                              & 230444    & 0.2 \\
 RAR compressed archive (v5.0)                                      & 192091    & 0.1 \\
 MinGW32 C/C++ Executable                                           & 175859    & 0.1 \\
 DirectShow filter                                                  & 156827    & 0.1 \\
 Windows ActiveX control                                            & 156373    & 0.1 \\
 Win32 EXE PECompact compressed (v2.x)                              & 153493    & 0.1 \\
 Win32 Executable Borland Delphi 6                                  & 147093    & 0.1 \\
 Win32 Executable PureBasic (generic)                               & 144066    & 0.1 \\
 Microsoft Input Method Editor                                      & 136382    & 0.1 \\
 Microsoft Excel sheet (alternate)                                  & 134536    & 0.1 \\
 ELF Executable and Linkable format (generic)                       & 133249    & 0.1 \\
 RAR compressed archive (v-4.x)                                     & 132273    & 0.1 \\
 Excel Microsoft Office Open XML Format document                    & 131081    & 0.1 \\
\hline
\end{tabular}
\caption{Top 50 TrID file types.}
\label{tab:trid_ftypes}
\end{table}

	\begin{figure}[]
		\includegraphics[scale=.5]{fig/all_exif_filetypes_pb.png}
		\caption{Top 20 file types from ExifTool output.
			\platon{Would be nice to split by platform. Would require some time to correctly
				group by platform. Integrate with trid information could help.}}
		\label{fig:exif_file_types}
	\end{figure}

	\begin{figure}[]
		\includegraphics[scale=.5]{fig/all_trid_filetypes_pb.png}
		\caption{Top 20 file types from Trid output. }
		\label{fig:trid_file_types}
	\end{figure}

	\begin{figure}[]
		\includegraphics[scale=.5]{fig/all_trid_exts_pb.png}
		\caption{Top 20 Trid file extensions. }
		\label{fig:trid_file_exts}
	\end{figure}

	Figure~\ref{fig:exif_file_types} shows the top 20 file types by number
	of samples identified by ExifTool and
	Table~\ref{tab:exif_ftypes} the top 50 file types.
		Figure~\ref{fig:trid_file_types} shows the top 20 file types by
	number of samples for Trid and
	Table~\ref{tab:trid_ftypes} the top 50 file types.

	\platon{I like more TriD granularity. However we need to some type of grouping.
		One dimension could be by platform (e.g., windows, linux, android, macos,
		multi-platform).}

}

\end{document}